\documentclass[preprint,12pt]{revtex4-1}

\usepackage{amssymb,amsmath,graphicx,graphics,amsfonts}
\usepackage[T1]{fontenc}
\usepackage[caption=false]{subfig}

\usepackage{libertine}

\newcommand{\U}{\mathbb{U}}
\newcommand{\Pe}{\mathbb{P}}
\newcommand{\A}{\mathbb{A}} 

\newcommand{\C}{\mathbb{C}}

\newcommand{\AT}{\mathbb{A}^{\footnotesize{\hbox{\scriptsize{T}}}}}

\newcommand{\F}{\mathbb{F}}

\newcommand{\V}{\mathbb{V}}

\newcommand{\Act}{\mathcal A}
\newcommand{\Aq}{\mathcal A_{\text{quad}}}
\newcommand{\btheta}{{\bf \theta}}
\newcommand{\Dd}{{\bf \textrm{D}}}
\newcommand{\dd}{{\bf \textrm{d}}}

\newcommand{\E}{\mathbb{E}}
\newcommand{\sigmaT}{\sigma^{\textrm{T}}}

\newcommand{\integ}{\int_{t_1} ^{t_2} {\bf \textrm{d}} t}
\newcommand{\integs}{\int_{s_1} ^{s_2} {\bf \textrm{d}} s}

\newcommand{\Tr}{\text{Tr}}
\newcommand{\be}{\begin{equation}} \newcommand{\ee}{\end{equation}} 
\newcommand{\ben}{\begin{eqnarray}}
\newcommand{\een}{\end{eqnarray}} 
\newcommand{\p}{\partial}

\newcommand{\lb}{\label}
\newcommand{\nn}{\nonumber}

\begin{document}

\title{Parametric Hamilton's equations for Lagrangian model for passive scalar gradients}
\author{L. S. Grigorio}
\affiliation{Centro Federal de Educa\c c\~ao Tecnol\'ogica Celso Suckow da Fonseca, Av. Governador Roberto Silveira 1900, Nova Friburgo, 28635-000, Brazil}

\begin{abstract}

In the context of instanton method for stochastic system this paper purposes a modification of the arclength parametrization of the Hamilton's equations allowing for an arbitrary instanton speed. The main results of the paper are: (i) it generalizes the parametrized Hamilton's equations to any speed required. (ii) corrects the parametric action on the occasion that the Hamiltonian is small but finite and how it adjusts to the probability density function (pdf). (iii) Improves instanton approximation to pdf by noise and propagator renormalization. As an application of the above set up we predict the statistics of passive scalar gradients in a Lagrangian model for turbulence, namely the scalar gradient Recent Fluid Deformation Closure (sgRFD).


\end{abstract}
\maketitle

\section{Introduction}


Turbulence is one the areas where observables such as velocity gradients and dissipation display huge fluctuations. The strong bursts in energy dissipation that take place at small scales, in turn, are responsible for the intermittency phenomenon, a long standing problem - breakdown of the Kolmogorov K41 theory. In face of that, instantons, known as the most probable path leading to a severe fluctuation, can be a valuable tool capable of  scrutinizing large events lying at small scales of turbulent flows shedding some light in our understanding of intermittency. For instance, applications of instanton approach  in turbulence-like phenomena appears in  \citep{Guraire_Migdal,Falkovic_etal_96,Chernykh_Stepanov}. See also \cite{Grafke_2015} and references therein for a numerical review in fluid dynamics.

In a broader context, the instanton approach is regarded as embedded in the framework of large deviation theory \cite{TOUCHETTE20091}, in which numerical or analytical evaluation of large fluctuation paths is a central question. When designing an efficient method to find such fluctuations one comes up against a problem resulting from how rare these trajectories arise. This fact turns prohibitive direct sampling and the need for alternative approaches comes in handy as those based on adaptive sampling or cloning trajectories. Some recent applications along these lines can be found in Refs. \cite{Bouchet_etal_2016_GL, Giardina_etal_PRL2006,Lestang_2018}. Another line of attack relies on action minimization by the so called minimum action method, where instantons play the role of minimizers \cite{mam,string,string2,pathways,gmam}. With respect to the action minimization, one important issue is that instanton speed approaches zero close to attractors and increases nearly exponentially as it departs from the attractors making duration of such trajectories infinite. To circumvent this Ref. \cite{arclength} purposes a set of arclength parametrized Hamilton's equations built on the reformulated geometric  action \cite{gmam}, such that the  instanton is endowed with a constant speed along the trajectory.

In this paper we generalize the geometric oriented parametrization of \cite{arclength} to account for arbitrary instanton speed. We also show how the parametric Hamilton's equations are derived from the Jacobi's/Mauperthuis' principle \cite{Lanczos}. This Jacobi/Mauperthuis action is equivalent to the geometric action of \cite{pathways,arclength,gmam}. In spite of both variational problems (Hamilton's principle and Jacobi's/Mauperthuis principle) lead to the same minimizer they may differ when the energy (Hamiltonian) does not vanish. This difference between the actions should be taken into account when estimating the transition probability density functions (pdfs). To illustrate, these ideas are apllied to the Lagrangian model for passive scalar gradients studied in \cite{gonzalez,rfd_magnetic}. We work out the statistics of the model as predicted by the instanton at different levels of approximation for small noise. A further improvement to the pdfs is obtained paying regard to instanton fluctuations by means of renormalization perturbation theory following \cite{Moriconi_2014,APOLINARIO}. The predicted statistics is compared with the statistics computed from solutions of stochastic equations.

The remainder of the article is organized as follows. Section II derives the parametric  Hamilton's equations along with the corresponding parametrized action. In section III the developed parametric equations  are applied to a model describing the dynamics of a Lagrangian passive scalar gradient in order to obtain the pdf. In addition to that, noise and propagator renormalization is performed to account for fluctuations around the optimal path. Finally, section IV is reserved for summary and perspectives.

\section{Parametric Hamilton's equations}
In this section we show how Hamilton's equations can be reparametrized in such manner that the instanton speed may be chosen arbitrarily. This is worked out by adapting the variational problem to restrict the search for paths with constant energy (Hamiltonian). The rationale consists in mapping the time evolution in a parametric representation of motion, eliminating the time dependence following closely reference \cite{Lanczos}.  To start with,  consider a vector $x \in \mathbb{R}^n$ obeying  the stochastic differential equation,
\be
\label{repar1}
\dot x = b(x) + \sigma \eta \, ,
\ee
where $\eta$ is a gaussian distributed random variable with zero mean and standard deviation equals to unity, and dot represents differentiation w.r.t. time $t$. We denote $D = \sigma \sigmaT $ as the diffusion matrix, assumed to be invertible, and $b(x)$ the drift term. In the path integral representation the action associated with the stochastic differential equation above is written as
\be
\label{repar2}
\Act = \integ \; L(x,\dot x),
\ee
with the Lagrangian,
\be
\label{repar2a}
L(x,\dot x) = \frac{1}{2} \left\langle \left(\dot{x}-b \right), D^{-1} (\dot{x} - b) \right\rangle \, ,
\ee
where $\langle x,y\rangle$ denotes the scalar product between vectors $x$ and $y$. Note that the noise correlator $D$ induces a metric such that $|\dot x|^2 _D \equiv \langle x,D^{-1}x \rangle $ so that the  Lagrangian can be written as $L = |\dot x - b |^2 _D /2$. 
As one is mostly interested in the stationary distribution of (\ref{repar1}), the time interval $t_2 - t_1$ should be taken to infinite. As pointed out in references \cite{gmam,arclength}, this leads to computational issues, since any discretization needs a finite time interval. Hamilton's equations calculated with the usual time parametrization eventually yields inaccurate instanton solutions. For instance, in the study of the stochastic Burgers equation, instantons could not be computed without a suitable time parametrization \citep{Chernykh_Stepanov,Grafke_2015,arclength}. Another reason to consider an alternative  parametrization is based on the fact that numerical evaluation performed with regular time spends much computation near the attractors, where instanton speed is quite slow, hence a more uniform instanton velocity turns the computation more efficient.

Consider a Lagrangian that does not depend explicitly on time and let us treat time as a dependent variable. This way, $x$ and $t$ will be regarded as functions of some parameter $s$ leading to a system of $n+1$ d.o.f.. The action take the form
\be
\label{repar2b}
\Act = \integs \; L \left(x,\frac{x'}{t'} \right) t' \, ,
\ee
where the prime denotes differentiation w.r.t. the independent variable $s$. Since $t$ now belongs to the set of dependent variables, its conjugated momenta can be defined ordinarily as 
\begin{equation}
\label{momenta}
p_t = \frac{\p (L t')}{\p t'} = L - \langle p, \dot x \rangle = -H \, .
\end{equation}

That is, the conjugated momentum corresponds to the negative of the Hamiltonian. The new Lagrangian in Eqn. (\ref{repar2b}) $L(x,\frac{x'}{t'})\,t'$ does not depend explicitly on $t$ from which it follows that the momentum associated with time is constant, $p_t =\text{const.}= -H = -E  $ (surrogate for Noether's theorem).  
We can make use of this constraint to eliminate $t'$ from the variational problem \cite{Lanczos} , first writing the reduced action $\bar \Act$ obtained by subtracting $p_t t'$ from the action \footnote{Can be interpreted as a Legendre transform.} 
\begin{align}
\label{Jacobi}
\bar \Act = \integs \left( Lt' - p_t t' \right)= \integs \;  \langle p , x' \rangle \, .
\end{align}
In this form, the variational problem performed on $\bar \Act$ is understood as a minimization   subject to the constraint $H = \text{const.}= E $. To complete $t'$ elimination we can make use of conservation of the Hamiltonian writing,
\begin{align}
\label{hamiltonian}
H =&  E = \frac{1}{2} \langle p ,D p \rangle + \langle p,b \rangle \, \\
\label{square}
E = & \frac 1 2 \,\langle (p + D^{-1} b) , D  (p + D^{-1} b) \rangle - \frac 1 2 \langle b, D^{-1} b \rangle \\
\label{conserv}
	= & \langle\dot x ,D^{-1} \dot x \rangle - \frac 1 2 \langle b, D^{-1} b \rangle \,  \Rightarrow | \dot x |^2 _D = 2 E + |b|^2 _D \, ,
\end{align}
where square completion was performed from (\ref{hamiltonian}) to (\ref{square}) and the relation $p = D^{-1}(\dot x - b)$ to go from (\ref{square}) to (\ref{conserv}). On account of the metric induced by the  diffusion matrix $D^{-1}$  one can define the line element $dl$ by the relation $d l^2 = \langle \dd x, D^{-1}  \dd x \rangle$. As a result, the last of Eqns. (\ref{conserv}) has a clear interpretation. The point $x(t)$ moves along a trajectory (instanton) with a speed $(dl/dt) = \sqrt{(2E + |b|^2 _D)}$, whereas in terms of the parameter $s$ one can write, in turn,
\begin{align}
\label{repar6}
\left(\frac{dl}{dt}\right)^2  = (2E + |b|^2 _D)\Rightarrow 
\left(\frac{dl}{ds}\right)^2 = (t')^2 (2E + |b|^2 _D) \, ,
\end{align}
which gives the rate of change of the local arclength with respect to the parameter $s$. Hence, choosing the parametrization accordingly allows us to dictate the velocity at which the point $x(s)$ runs the phase space.
After substituting (\ref{repar6}) in (\ref{Jacobi}) together with the definition of the momentum we arrive at
\begin{equation}
\label{Abar}
\bar \Act = \integs \; \left( \frac{dl}{ds} \sqrt{2E + |b|^2_D} - \langle x',D^{-1}b \rangle\right) \, .
\end{equation}

Minimizing the functional  (\ref{Abar}) in place of (\ref{repar2}) is known as the Jacobi's/Mauperthuis' principle. An alternative  variational problem which searches curves in configuration space  constrained to constant Hamiltonian \cite{Scholarpedia_leastaction}. By contrast, the variational problem concerning the action $\Act$ (Hamilton's principle) does not search for paths obeying this constraint $H = E$. Though, at end of the process, the minimizer of $\Act$ has constant Hamiltonian so that both minimizers agree.  

Since time does not figure in $\bar \Act$ the information about how the system evolves in time is lost. The instanton now is a path resting on configuration space, even though the time information can be retrieved after (\ref{repar6}). 

We underscore that by setting $E = 0$ and choosing the parametrization such that the speed is constant, \emph{i.e.}, $dl/ds = |x'|_D = \text{const.}$ we recover the geometric action obtained in \cite{gmam} and \cite{arclength}. Hence the action (\ref{Abar}) generalizes the aforementioned result allowing for any instanton speed by simply defining $dl/ds$ accordingly. A straightforward calculation shows the Hamilton's equations obtained from the parametric action (\ref{Abar}) take the form, 
\begin{align}
\label{eqs1}
&  x'(s) = \frac{dl}{ds}  \, \frac{b + D p}{\sqrt{2E + |b|^2_D}} \\ 
\label{eqs2}
&  p'(s) = - \frac{dl}{ds}\, \frac{\nabla b \, p}{\sqrt{2E + |b|^2_D}} \, ,
\end{align}
subject to suitable boundary conditions, where the operator $\nabla$ stands for differentiation with respect to the vector $x$, such that $(\nabla b)_{ij} = \p_i b_j = \p b_j/\p x_i$. The parametrization can be chosen freely by determining the speed at which  $x(s)$ goes through  the curve. Again, as long as $dl/ds$ is set to a constant and $E$ set to zero (\ref{eqs1}) and (\ref{eqs2}) match the arclength Hamilton's Eqns. proposed in Ref. \cite{arclength}.

However, we underline that $\Act$ and $\bar \Act$ in general differ even though their minimizers coincide. From (\ref{Jacobi}) and (\ref{momenta}) we have
\be
\label{A_barA}
\bar \Act = \integs \;t' (L + H) = \Act + \integ H = \Act + E(t_2-t_1) \, ,
\ee
where hamiltonian conservation was used. It is clear that $\bar \Act$ and $\Act$ agree only when $E$ vanishes, and indeed, this is required if one is interested in infinite long trajectories ($t_2 - t_1 \rightarrow \infty$). Though, during calculation of Hamilton's equations (\ref{eqs1}) and (\ref{eqs2}) numerically, the limit $E=0$ is delicate when transition envolves the critical point where $b$ vanishes. To prevent singularities it is suitable to consider a small but finite $E$. Consequently, when estimating the transition probability  this difference should be regarded since the pdf in the instanton approximation is given by $\rho \approx \exp(-\Act_{\text{min}}) = \exp(-\bar \Act_{\text{min}} + E \Delta t)$, where $\Act_{\text{min}}$ and $\bar \Act_{\text{min}}$ are the actions taken on the instanton trajectory. The point is, we can use the Jacobi's/Mauperthuis' principle taking advantage of geometric properties of $\bar \Act$, whose minimization leads to (\ref{eqs1}) and (\ref{eqs2}). But, as we shall show in an example, evaluating $\rho$ using $\Act$ yields a better estimation of pdf compared to the estimation with $\bar \Act$. Therefore, we are led to calculate the correction term $E \Delta t$ in Eqn.(\ref{A_barA}). A simple way to accomplish this is to use (\ref{repar6}) combined with (\ref{A_barA}) yielding
\begin{align}
\label{A_corr}
\Act = \bar \Act - \integs \; t' E \; \Rightarrow \; 
\Act = \bar \Act  - \integs \; \frac{|x'|_D\, E}{\sqrt{(2E+|b|_D^2)}} \, .
\end{align}
Finally, after (\ref{Abar}) we may write the action as
\begin{equation}
\label{A}
\Act = \integs \left[ \frac{(E+|b|_D^2)}{\sqrt{(2E+|b|^2_D)}} |x'|_D- \langle x',D^{-1} b \rangle \right] \, ,
\end{equation}
which turns out to be the original action put into a parametric form. An alternative way to arrive at Eqn. (\ref{A}) consists in the following. Departing from (\ref{Abar}) we implement energy conservation $E = (|x'|^2_D/(t')^2 - |b|^2_D)/2$ as a constraint with the help of a Lagrange multiplier and further solving for $t'$ to obtain (\ref{A}). In such manner the two variational problems are reconciled. Note that the parametric action (\ref{A}) is the same as  
the original time parametrized action (\ref{repar2}), yet rewritten in a time independent geometric form. Therefore, (\ref{A}) allows us to take advantage of the geometric parametrization, which in turn allows for a control of instanton speed, and at the same time correcting finite energy $E$ and finite time duration effects. 


\section{Application: The Lagrangian Recent Fluid Deformation for passive scalar gradient}

As an application of the formalism outlined above let us consider a model developed for fluctuations of the gradient of passive contaminant in Lagrangian turbulence \cite{gonzalez}. Our goal is to evaluate the stationary pdf as predicted by the instanton approximation taking into account small  instanton fluctuations. 

\subsection{Model equations}

Some Lagrangian closures have been conceived as a way to model the small scales of turbulence. Most of them rely on the Restricted Euler closure which neglects viscosity and suffers from finite time singularity. Among the models proposed to regularize this singularity there is the Recent Fluid Deformation (RFD), which fall back on  assumptions that the Cauchy-Green deformation tensor has a short time memory \cite{RFD}. As an  extension to the this model Refs. 
\cite{gonzalez} and \cite{rfd_magnetic} have considered the dynamics of the gradient of a passive scalar $\theta$ within the framework of the RFD, given by the following SDE 
\begin{align}
\lb{1}
\dot \theta_i = - A_{ik}\, \theta_k  - \frac{\Tr (\C^{-1})}{3T_\theta} \theta_i + g_\theta f_i \, ,
\end{align}
which we call scalar gradient Recent Fluid Deformation (sgRFD) closure,
where $\theta_i$ is understood as a shorthand for  the gradient of the passive scalar, $\theta_i = \p_i \theta$, and the dot denotes material time derivative $d/dt$. Latin indices  run from 1 to 3. The passive admixture gradient is coupled to the dynamics of the velocity gradient $\A$ whose components are $A_{ij} = \p_i u_j$, and $u_i$ is the velocity of the Lagrangian particle along direction $i$. The first term of the r.h.s. of (\ref{1}) stems from the advective transport and second term models diffusion by approximating the Cauchy-Green tensor $\C$ as $\C^{-1} = \exp[-\tau \A] \exp[-\tau \A^{\text{T}}]$ with $\tau$ the Kolmogorov time scale. The gradient scalar dynamics is supplemented with a white noise force $f_i$ such that correlations read $\E[f_i(t) f_j(t')] = \delta_{ij} \, \delta(t-t')$ and $g_\theta$ is the noise amplitude. The parameter  $T_\theta$ is a time scale related to Lagrangian diffusion time $\approx \lambda^2/\kappa$, being $\kappa$ the diffusive constant and $\lambda$ the Taylor microscale.
 
The sgRFD equation (\ref{1}) depends on the evolution of the velocity gradient $A_{ij}$ which within the RFD \cite{RFD} closure reads,
\be
\lb{4}
\dot \A =  - \A^2 +  \frac{\C^{-1} \Tr( \A^2) }{\Tr(\C^{-1})} - \frac{\Tr(\C^{-1})}{3T} \A  + g \mathbb{F} \, .
\ee
The first term on the right hand side is exact and relate to self-stretching, the second one models the non-local interactions stemming from the pressure gradient and the third term is a tantamount to the viscous dissipation, where $T$ denotes a time scale of order $\lambda^2/\nu$ with $\nu$ the kinematic viscosity. Added to the deterministic part of the dynamics there is a zero average white noise tensor  $\F$ with amplitude $g$ whose correlation function is prescribed by
\be
\lb{5}
\E \left[F_{ij}(t) F_{kl}(t')\right]= G_{ijkl}  \delta(t-t') \ ,
\ee
with
\be
\lb{6}
G_{ijkl}=2 \delta_{ik} \delta_{jl} - \frac{1}{2} \delta_{il} \delta_{jk}- \frac{1}{2} \delta_{ij} \delta_{kl} \ . \
\ee

In the remainder of the paper we set $T$ to unity, $T_\theta=0.5$, $\tau = 0.1$ and $g_\theta = 1$. This choice of parameters corresponds to flows with Schmidt number close to unity \cite{gonzalez} and modest Reynolds number around 10. Another reason is practical one since as pointed out in \cite{gonzalez}, also confirmed by our simulations, higher values of $T_\theta$ (closer to or higher than unity) may lead to uncontrolled  instabilities during stochastic integrations as had been tested using both stochastic recipes \cite{Honeycutt_92} and \cite{WELTON1997}.

\subsection{Path integral}

Consider  that at $t = -\infty$ the passive scalar gradient vanishes and as a result of fluctuations due to external forcing and coupling to velocity gradient, the component $\theta_1$ (without loss of generality since the model is isotropic) takes on a value $k$ at time $t=0$. The probability of such transition can be written with the help of the  Martin-Siggia-Rose/Janssen/de Dominicis  path integral  \cite{PhysRevA.8.423,Janssen1976,Dominicis_76} like,
\be
\lb{path_int}
\rho(k) =  \int_{\{\Sigma\}}  \Dd [\Pe] \Dd [\A] \Dd [\Pi] \Dd [\theta] \; \exp \left\{\begin{matrix}
-\Act[\Pe, \A, \Pi, \theta ]
\end{matrix}   \right\}  \, ,
\ee
subject to the set of boundary conditions $\Sigma$, 
\be
\label{boundary}
\Sigma = \left\{ \A(-\infty) = 0 \, ,\, \theta_j(-\infty) = 0\, , \, j=1,\dots,3 \;\;\text{and} \;\; \theta_1(0) = k \right \} \, .
\ee 
In (\ref{path_int}) the action functional reads 
\be
\lb{action}
\Act[\Pe, \A, \Pi, {\bf \theta}] =  \int_{-\infty}^{0} \text{d}t \, \left[\text{Tr} \left( \Pe^{\text{T}} (\dot \A - \V[\A]) \right)- \frac{g^2}{2} P _{ij} G_{ijkl} P_{kl}  + \Pi_j (\dot \theta_j - M_j [\theta,\A]) - \frac{g_\theta ^2}{2} \Pi_i \Pi_i\right] \, ,
\ee
where 
\begin{equation}
\label{dettheta}
M_i[\theta,\A] = - A_{ij}\, \theta_j  - \frac{\Tr (\C^{-1})}{3T_\theta} \theta_i \, ,
\end{equation}
and
\begin{equation}
\label{detA}
\V[\A]= - \A^2 +  \frac{\C^{-1} \Tr( \A^2) }{\Tr(\C^{-1})} - \frac{\Tr(\C^{-1})}{3T} \A \, ,
\end{equation}
stand for the deterministic part of (\ref{1}) and (\ref{4}), respectively, and $\Pe$ and $\Pi$ are the auxiliary variables assigned to $\A$ and $\theta_i$. 

We aim at evaluating the most probable trajectory related to the transition $\Sigma$. 
Computation of the referred trajectory can be achieved via minimization of the action functional (\ref{action}). The knowledge of this optimal path allows us to estimate the transition pdf according to the saddle point approximation as $\rho(k) \approx \exp(-\text{min}_{\{\A , \theta , \Pe , \Pi\}}[\Act])$ subject to the conditions $\Sigma$ (\ref{boundary}).

Before presenting the results we calculate how fluctuations around instanton path contribute to the pdf's according to renormalization techniques in next subsection.


\subsection{Infrared noise renormalization}
The most probable path is the major contribution to the transition probability (\ref{path_int}) and in the limit of small noise amplitude it is sufficient to describe the statistics. However, for finite noise amplitude it is necessary to consider  fluctuations around instanton trajectory. These fluctuations may be accounted for by the renormalization procedure \cite{mccomb}, \cite{barabasi-stanley}. To this end, we calculate noise renormalization following closely the rationale presented in reference \cite{Moriconi_2014} and propagator renormalization according to reference \citep{APOLINARIO}, which also provides a detailed analysis of perturbation techniques applied to the RFD including an estimation of Feynman diagrams at higher orders.

We start by rewriting the action in the original time parametrization (\ref{action}) in a way that bilinear terms in $\Pi$, $\Pe$, $\A$ and $\theta_i$ are written explicity, after expanding $\Tr \, (\C^{-1})$ in (\ref{detA}) and (\ref{dettheta}) up to order $\mathcal O (\tau^2)$,
\begin{align}
\lb{act}
\Act  = \int_{-\infty}^{0} \text{d} t \left[ \Pi_k (\dot \theta_k + \theta_k /T_\theta) + \Pi_i N_{ij} \theta_j - \frac{g_\theta ^2}{2} \Pi_i \Pi_i + \right. \nn \\
  \left. + \Tr \left( \Pe^{\text{T}} (\dot \A + \A/T) \right) - \frac{g^2}{2} P _{ij} G_{ijkl} P_{kl} - \Tr \left(\Pe^{\text{T}} \U \right) \right]\, ,
\end{align}
where $N_{ij}$ represents the coupling between the dynamics of $\btheta$ and $\A$, namely
\be
\label{N}
N_{ij}(\A) = A_{ij} - \frac{2\tau}{3T_\theta} \text{Tr} \A \, \delta_{ij} + \frac{\tau^2}{3T_\theta} \Tr \left( \A^2 +  \AT \A \right) \delta_{ij} \, ,
\ee
and $\U(\A)$ contains non linear terms in $\A$ present in $\V(\A)$, that is,
\be
\lb{U}
\U(\A) = \V(\A) + \A/T \, . 
\ee
One can define the free action and quadratic actions by
\begin{align}
\label{A0}
\Act _0 =  & \int \dd t \, \left[ P_{ij}(t) (D^{-1}_{ijkl} * A_{kl} )(t) + \Pi_i(t) (R^{-1}_{ij}*\theta_j)(t)\right] \, , \\
& \Aq = \Act_0 - \int \dd t \left[ \frac{g_\theta^2}{2} \Pi_i \Pi_i + \frac{g^2}{2} P_{ij}G_{ijkl}P_{kl}\right] \, .
\end{align}
where $*$ stands for the convolution and we have the operators \cite{Chow2010PathIM},
\be
 D^{-1}_{ijkl}(t-t') =\delta_{ik} \delta_{jl}\,  \delta(t-t') (\p_{t'}  + 1/T) \quad \text{and} \quad R^{-1}_{ij}(t-t') = \delta_{ij}\delta(t-t') (\p_{t'}  + 1/T_\theta) \, .
\ee  
Now, (\ref{act}) can be written as $\Act = \Aq +  \Act_I$, with the interaction term,
\begin{align}
\lb{deltaS}
\Act_I [\Pe, \A, \Pi, {\bf \theta}]= \int_{-\infty}^0 \dd t \left[ \Pi_i N_{ij}(\A) \theta_j - \Tr \left( \Pe^{\text{T}} \U(\A) \right)\right] \, .
\end{align}


Expectation values w.r.t the free action $\Act _0$ are calculated with the expression
\begin{equation}
\label{free_expec}
\E_0[\dots] = \int \Dd [\Pe] \Dd [\A] \Dd [\Pi] \Dd [\theta] \, [\dots]\exp \left\{ \begin{matrix}
-\Act _0
\end{matrix} \right\}  \, .
\end{equation} 

\begin{figure}
\includegraphics[scale=.95]{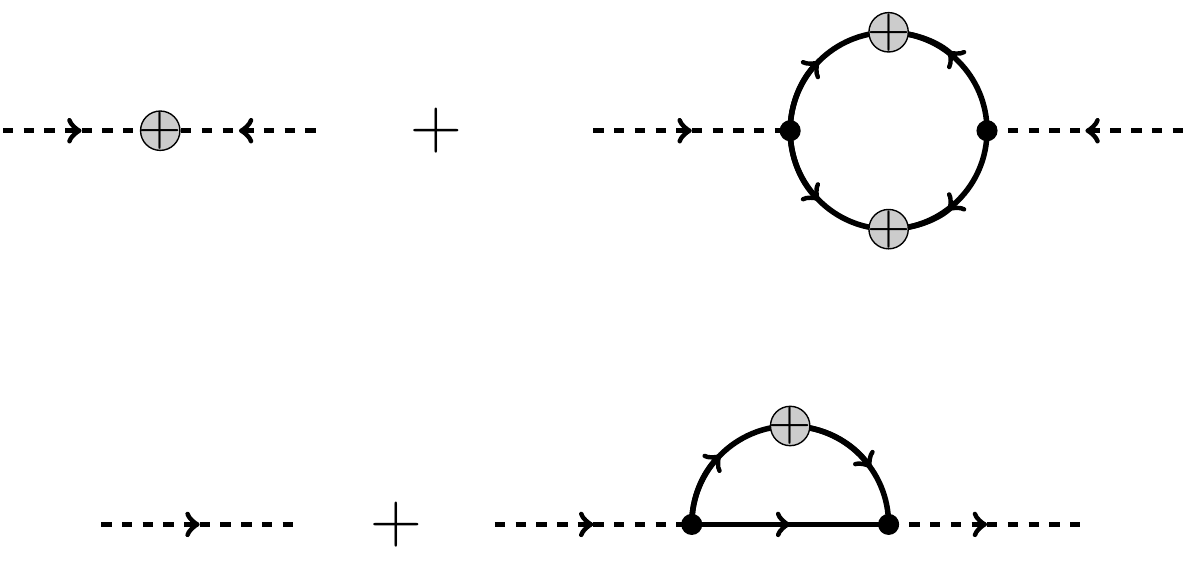}
\caption{One-loop Feynman diagrams accounting for: Top: noise renormalization. Bottom: Propagator renormalization. Black filled circles correspond to the vertices. Likewise the crosses (+) inside gray filled circles amount to noise terms. Arrows indicate the wave number direction. Each solid line represents the two point functions $R_{ij}$ or $D_{ijkl}$ (\ref{twopoint}).} 
\label{feynman}
\end{figure}

 The so-called free two point functions can be shown to be given by \cite{APOLINARIO},
\begin{align}
\label{twopoint}
& \hspace{2cm} \E_0 [A_{ij}(t) P_{kl}(t')] = D_{ijkl}(t-t')\, , \; \; \E_0[\theta_i(t) \Pi_j(t')] = R_{ij}(t-t'). \nn \\
& D_{ijkl}(t-t') =  H_e(t-t')\, \delta_{ik}\delta_{jl}\, e^{-\frac{(t-t')}{T}} \;\;\;  \text{and} \;\;\; R_{ij}(t-t') = \delta_{ij}\, H_e(t-t') \,e^{-\frac{(t-t')}{T_\theta}}\, ,
\end{align}
where $H_e(t)$ is the Heaviside function with $H_e(0) = 0$ (consistent with It\^o convention).  These two point correlation functions are the bulding blocks for the perturbative expansion.
As a further step we split the dynamical variables into a sum of fast and slow modes, in accordance with standard techniques (see for instance \cite{mccomb,barabasi-stanley}), 
\begin{align}
& \A = \A^{+} + \A^{-},  \quad {\bf \theta} = {\bf \theta} ^{+} + {\bf \theta}^{-} \\
 & \Pe = \Pe^{+} + \Pe^{-}, \quad \Pi = \Pi^{+} + \Pi^{-} \, ,
\end{align}
where $\A^{+}$ ($\A^{-}$) and conversely $\theta_i^{+}$ ($\theta^{-}_i$), denotes the sum over modes $\omega$ above (below) the threshold $\omega_c$ such that,
\begin{align}
\label{modes}
\A^{+} = \int_{\{ |\omega| > \omega_c \}} \hspace{-0.7cm} \dd \omega \;\; e^{i \, \omega \,t} \hat{\A}(\omega)/2\pi \, ,\qquad \A^{-} = \int_{\{ |\omega| < \omega_c \}} \hspace{-0.7cm} \dd \omega \;\; e^{i \, \omega \,t} \hat{\A}(\omega)/2\pi \; , \\
\theta_i ^{+} = \int_{\{ |\omega| > \omega_c \}} \hspace{-0.7cm} \dd \omega \;\; e^{i \, \omega \,t} \hat{\theta_i}(\omega)/2\pi \, ,\qquad \theta_i^{-} = \int_{\{ |\omega| < \omega_c \}} \hspace{-0.7cm} \dd \omega \;\; e^{i \, \omega \,t} \hat{\theta_i}(\omega)/2\pi \; , 
\end{align}
where $\hat{\A}(\omega)$ and $\hat \theta_i (\omega)$ are the Fourier transform of $\A(t)$ and $\theta_i(t)$, respectively.
The next step consists of eliminating the fast variables in favor of the slow ones by summing these fast modes. This allows one to define the effective action \citep{mccomb,Moriconi_2014}
\begin{align}
\lb{eff_act1}
\exp\left\{ \begin{matrix} -\Act_{\text{eff}}[\Pe, \A, \Pi, {\bf \theta}] \end{matrix}\right\} & =  e^{-\Act[\Pe^{-}, \A^{-}, \Pi^{-}, {\bf \theta^{-}}]} \int  \Dd [\Pe^{+}] \Dd [\A^{+}] \Dd [\Pi^{+}] \Dd [\theta^{+}] \exp\left\{\begin{matrix} -\Act^+ -\Delta \Act \end{matrix} \right\} \, \nn \\
&= \exp\left\{ \begin{matrix} -\Act[\Pe^{-}, \A^{-}, \Pi^{-}, {\bf \theta^{-}}] \end{matrix} \right\} \, \E \left[ e^{-\Delta \Act}\right] \, .
\end{align}
In equation (\ref{eff_act1}) $\Act^+$ represents (\ref{act}) evaluated with fast variables, and
\begin{align}
\lb{deltaS}
\Delta \Act = \int_{-\infty}^0 \dd t\left\{ (\Pi^{+}_i   \theta^{-}_j +\Pi^{-}_i \theta^{+}_j)(N_{ij}(\A^+)+ N_{ij}(\A^-)) +\Pi^{+}_i N_{ij}(\A^-)\theta^{+}_j+ \right.\nn \\
+ \left. \Pi^{-}_i  N_{ij}(\A^+) \theta^{-}_j-\Tr \left[ (\Pe^+) ^{\text{T}} (\U(\A^-)+\Delta \U) \right] -\Tr \left[ (\Pe^-)^{\text{T}} (\U(\A^+) +\Delta \U) \right] \right\} \, .
\end{align}
which mixes fast and slow variables through the coupling terms and,
\begin{equation}
\label{deltaU}
\Delta \U = \U(\A^+ + \A^-) - \U(\A^+) - \U(\A^-) \, .
\end{equation}
Among the many terms that contribute to the effective action (\ref{eff_act1}) we will keep those which yield the lowest order contribution in powers of $g$ (one loop) as depicted in the Feynman diagrams Fig. (\ref{feynman}). Note in passing that crossed terms which could arise from the nonlinear terms in (\ref{N}) does not contribute to the one loop perturbation, so they are not written in (\ref{deltaS}).

Noise renormalization stems from the term Fig. (\ref{feynman}) top,
\be
\lb{noise_nor}
\frac 1 2 \Pi^- _i \Pi^- _j \; \E [N_{ik}(\A^+) \theta_k^+ N_{jl}(\A^+) \theta_l^+] \, ,
\ee
whereas propagator renormalization, Fig. (\ref{feynman}) bottom, derives from the term,
\be
\lb{prop_nor}
\E [\Pi^- _i N_{ij}(\A^+)\theta_j \Pi^+ _k N_{kl}(\A^+)\theta^{-}_l] \, .
\ee
Straightforward computation yields
\be
\label{gtheta_ren}
\{g_\theta ^2\}^{\textsc{ren}} = g_\theta ^2  \left( 1 +  \frac 5 2 \, g^2 \,\frac{(T T_\theta)^2}{T+T_\theta} \right) \, .
\ee
Recall that in the following the parameters are set to $T=1$ and $T_\theta = 0.5$.
Similar evaluation results in vanishing contribution for propagator renormalization, that  is, Eqn. (\ref{prop_nor}) results in zero. Tensor contractions were computed with the help of \cite{MathGR}.

\subsection{Numerical implementation}

Action minimization can be speed up by decreasing the number of degrees of freedom if we exploit how symmetries come into play in (\ref{action}), in particular rotation symmetry. First, note that the variational problem consists of minimizing the action between initial point $(\A = 0, \,\theta=0$) and final point $\theta_1 = k$ ($k\neq 0$), with all other components of $\theta_i$ and of $\A$ unconstrained at the endpoint. Naturally, if no imposition is made at the endpoint, the minimization would be fulfilled by  $\A(t) = 0$ and $\theta_i(t) = 0$ along the entire trajectory leading also to a vanishing action. However, the final condition on $\theta$ prevents $\theta_1(t)$  from vanishing. Since the velocity gradient drives the fluctuation of the passive scalar it is reasonable that a non zero value of $\A$ is necessary to yield $\theta_1(0) = k$ from $\theta_1(-\infty)=0$. On the other hand, the final condition could be alternatively taken into account as an additional term in the action written in the form $\int \dd t \delta(t) (\theta_1(t) - k)$. The presence of this term breaks the rotation symmetry  except for rotation along the $x_1$ direction. Invariance under rotation along $x_1$ axis implies that the velocity gradient should be diagonal with $A_{22} = A_{33}$. One more consideration is the incompressibility condition which results in the following form acquired by the instanton velocity gradient $\A =\text{diag} (a(t),-a(t)/2,-a(t)/2)$. 

\begin{figure}[h!]
\centering
\subfloat[]{\includegraphics[scale=.9]{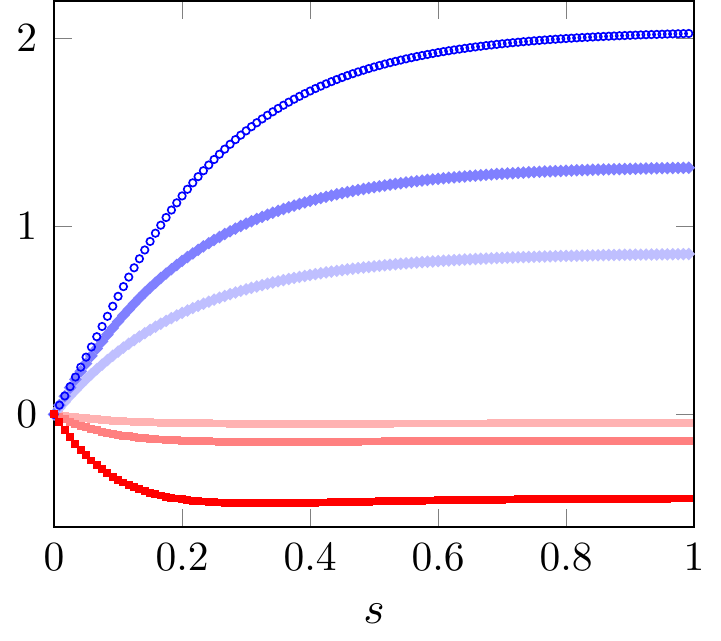}}
\subfloat[]{\includegraphics[scale=.9]{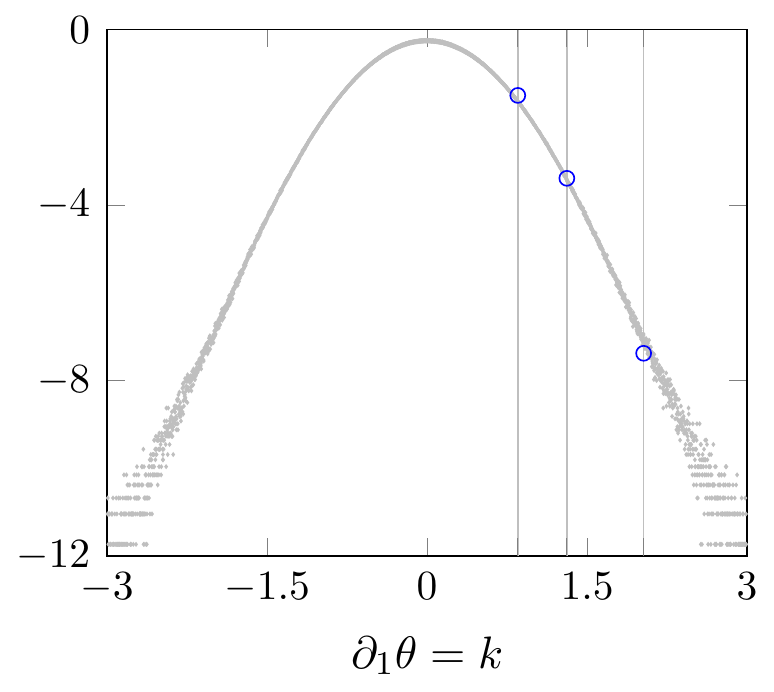}}
\caption{(a) Numerical instanton as solution of the parametric  Hamilton's equations (\ref{eqs1})-(\ref{eqs2}). Blue stands for $\theta_1(s)$ and red stands for $A_{11}(s)$. Lighter to darker correspond to $\theta_1(0)$ (scalar gradient final value) equals to $0.85$, $1.31$ and $2.03$, respectively. As can be seen in all three cases the velocity gradient $A_{11}$ is negative corresponding to a compressional direction. Parameters: $g=0.5$ and $E = 0.3$. (b) log of scalar gradient pdf derived from SDE's (\ref{1}) and (\ref{4}). Each cut is assigned to a solution in (a). The variance and kurtosis are 0.27 and 3.17, respectively.} 
\label{fig0}
\end{figure} 

Therefore, the number of d.o.f. left is reduced to two,  $\vartheta \equiv \theta_1$ and $a \equiv A_{11}$, in place of the 8 independent  velocity gradient variables plus 3 passive scalar gradient, markedly lowering the computational effort. This assertion can be confirmed {\it a posteriori} performing the minimization with the full set of variables. This computational cost saving with the use of symmetries have been presented in \cite{Grigorio_2017} in the context of RFD closure.

With the preceding considerations we are led to minimize the low dimensional action
\begin{equation}
\lb{ALD}
\Act _{\text{LD}} = \int_{-\infty}^0 \dd t \;\left[ p \left(\dot a -  b_1(a) \right) - \frac {g^2}{2} p^2 + \Pi\left(\dot \vartheta - b_2(\vartheta,a)\right) -\frac{g_\theta^2}{2} \, \Pi^2 \right]\, ,
\end{equation}
with $a$ and $\vartheta$  corresponding to $A_{11}$ and $\theta_1 = \p_1 \theta$ from the original action (\ref{action}), respectively and 
\begin{align}
\lb{determ} 
b_1(a) = -a^2 & + \frac 3 2 a^2 \frac{e^{-2\tau a}}{e^{-2 \tau a}+ 2 e^{\tau a}} - \frac{a}{3} (e^{-2 \tau a}+ 2 e^{\tau a}) \, , \\
b_2(\vartheta,a) &= -\vartheta a - (e^{-2 \tau a}+ 2 e^{\tau a}) \frac{\vartheta}{3T_\theta} \, ,
\end{align}
as the low dimension version of $\V$ (\ref{detA}) and $M$ (\ref{dettheta}).
In face of (\ref{A}) we can suitably define 
\begin{equation}
\label{Ax}
x = \left(\begin{array}{c}
a \\ 
\vartheta
\end{array} \right) \, ,  \quad D = \left(\begin{matrix}
g^2 & 0 \\ 
0 & g_{\theta}^2
\end{matrix} \right)\, , \quad b = \left(\begin{array}{c}
b_1 \\ 
b_2
\end{array} \right) \, ,
\end{equation} 
to rewrite (\ref{ALD}), after elimination of $p$  and $\Pi$, as 
\begin{equation}
\label{ALD2}
\Act_{\text{LD}} = \int_{-\infty}^0 \dd t \; \frac 12 |\dot x - b|_D^2 \, ,
\end{equation}
such that parametric Hamilton's equations  (\ref{eqs1}), (\ref{eqs2}) subject to set of conditions (\ref{boundary}) can readily be applied to find the minimizer of (\ref{ALD2}).  To solve them we adapt the routine prescribed in \cite{arclength}, which in turn is a variation of the  iterative Chernykh-Stepanov method \citep{Chernykh_Stepanov}, as follows.

\begin{enumerate}
\item Choose a function $f(s)$ which will dictate instanton speed as $\dfrac{dl}{ds} = q f(s)$. $q$ is a constant to be adjusted in step 4.
\item Propagate backwards in parameter $s$ equation $(\ref{eqs2})$ for $p(s)$ with a chosen final condition $p(1)$ for a given approximation of $x(s)$ and $q$.
\item Propagate forwards in parameter $s$ equation (\ref{eqs1}) for $x(s)$ with the previously computed $p(s)$ and $q$.
\item Compute the arclength $L = \displaystyle{ \integs }\,  |x'(s)|_D$ from step 3. Adjust instanton speed by setting the constant to $q = \dfrac{L}{\displaystyle{\integs } f(s)}$.
\item Repeat steps from (2) to (4) until convergence is fulfilled. 
\end{enumerate}

 During iterative steps, the Hamiltonian  converges to the chosen value $E$. Likewise, $q$ converges towards a fixpoint. We recall that a non vanishing $E$ prevents the solution to blow up.  Figure (\ref{fig0}a) displays three instantons evaluated according to the prescription above for $g=0.5$, $g_\theta = 1.0$ and $E=0.3$. Blue circles corresponds to scalar gradient $\vartheta$ while red squares represents the longitudinal velocity gradient $a$. $s$ is taken in the interval $[0,1]$ divided in $N=120$ steps with integration performed by Heun method. Given the relative deviation $\delta = |\vartheta(1)-\vartheta^{\textsc{prev}}(1)|/\vartheta^{\textsc{prev}}(1)$, where $\vartheta^{\textsc{prev}}$ denotes the previous iteration, we adopted $\delta<0.5\times10^{-5}$ as criteria for convergence. Fig.(\ref{fig0}b) exhibits the log of numerical pdf derived from SDEs (\ref{1}) and (\ref{4}) and the log of the pdf predicted according to each instanton in the left Fig.(\ref{fig0}a). More specifically, open blue symbols are the values of the negative action on each of instanton trajectory (\ref{A}) depicted in Fig. (\ref{fig0}a).     
 
\begin{figure}[h]
\centering
\subfloat[]{\includegraphics[scale=.9]{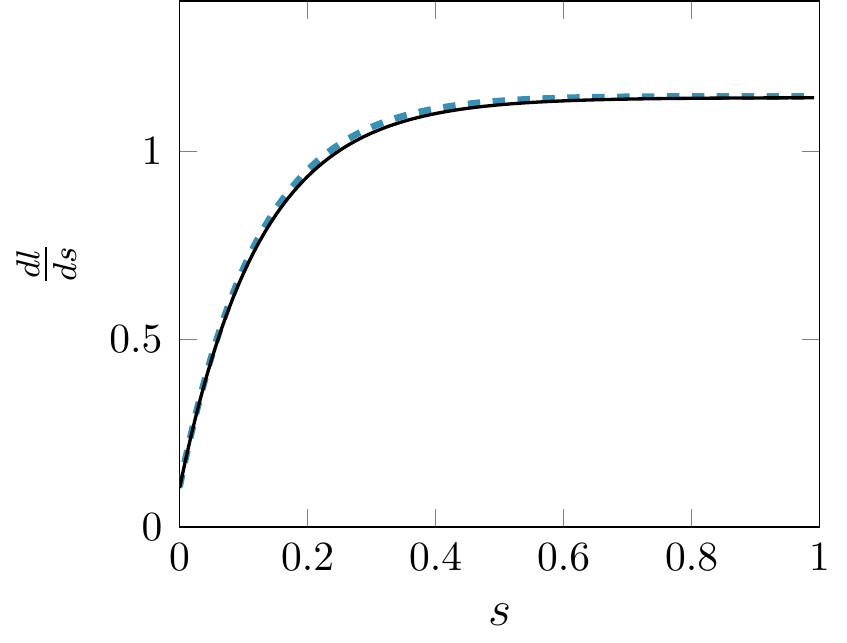}} 
\subfloat[]{\includegraphics[scale=.9]{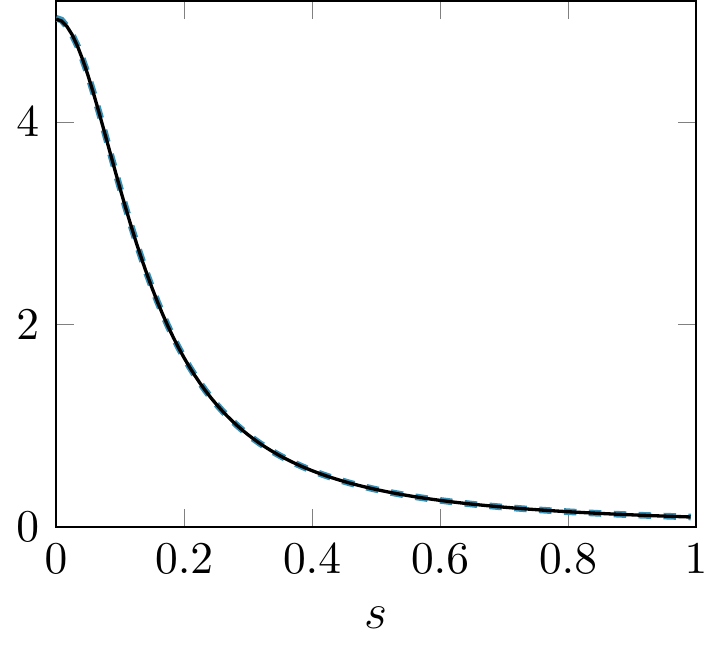}}
\caption{Plot of $dl/ds = qf(s)$ used as input (solid line) compared with numerical instanton speed $|x'|_D$ after convergence (dashed) as functions of parameter $s$. Both evaluated for same parameters of Fig. (\ref{fig0}). (a) $dl/ds = 1.04\,[1.1-\exp(-8s)]$. (b) $dl/ds = 0.10/(0.02+s^2)$. Agreement between the curves is established only after convergence is reached.} 
\label{fig1}
\end{figure}

In the course of iterations, $|x'(s)|_D$ approaches the input speed $dl/ds=qf(s)$. Two instances of instanton speed as a function of $s$ is shown in Fig. (\ref{fig1}), more specifically, a comparison between numerical $|x'(s)|_D$ after convergence (dashed), and input $dl/ds$ (solid line). Fig. (\ref{fig1}a) plots $dl/ds = 1.04\,[1.1-\exp(-8s)]$ while (\ref{fig1}b)  plots $dl/ds = 0.10/(0.02+s^2)$.  

\begin{figure}[h]
\centering
\subfloat[]{\includegraphics[scale=.9]{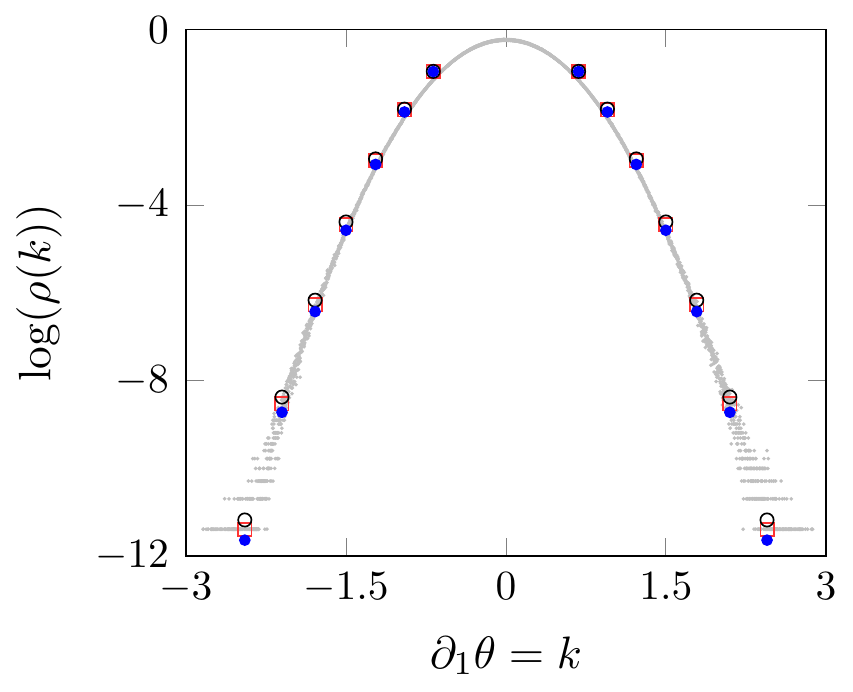}} 
\subfloat[]{\includegraphics[scale=.9]{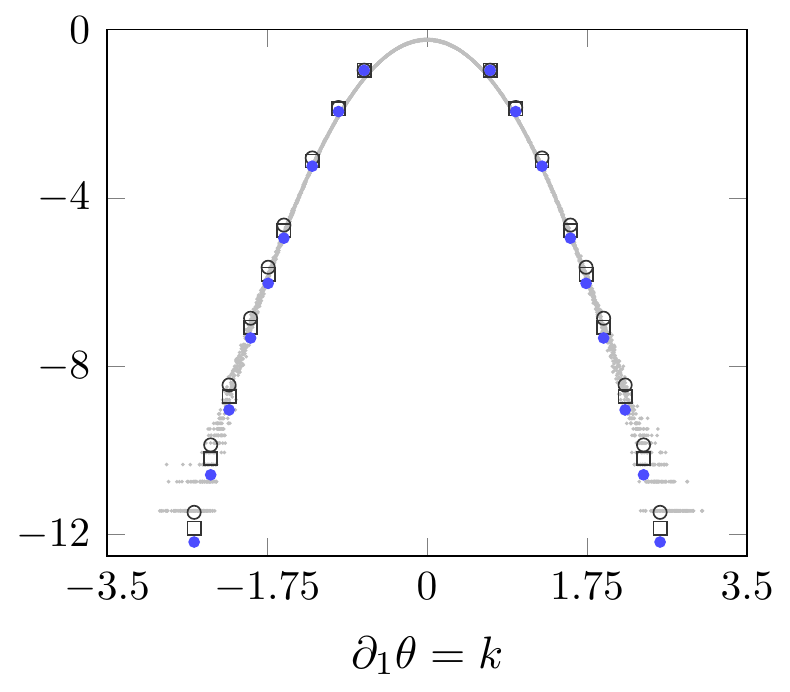}} \\
\subfloat[]{\includegraphics[scale=.9]{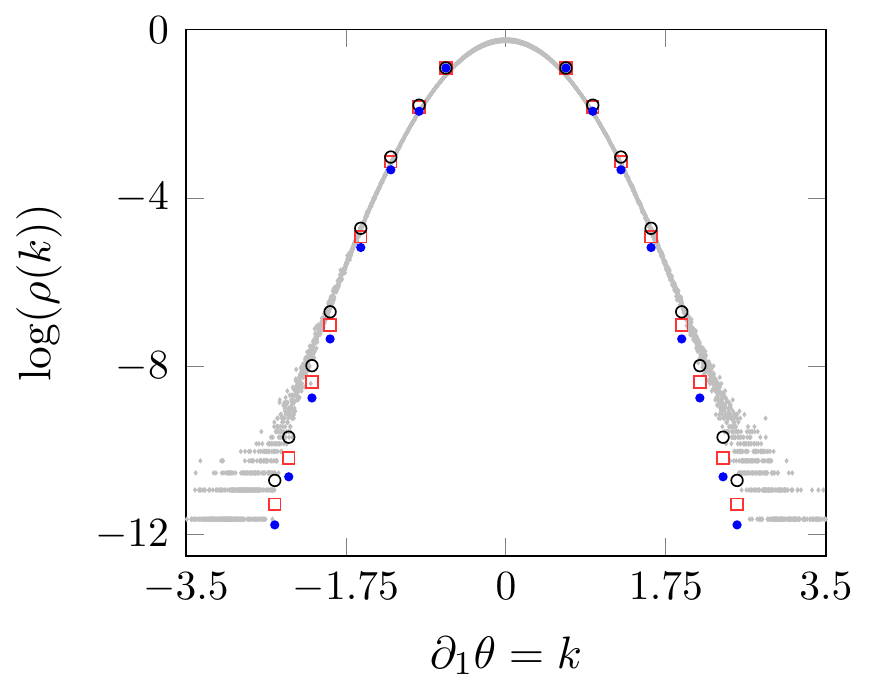}}
\subfloat[]{\includegraphics[scale=.9]{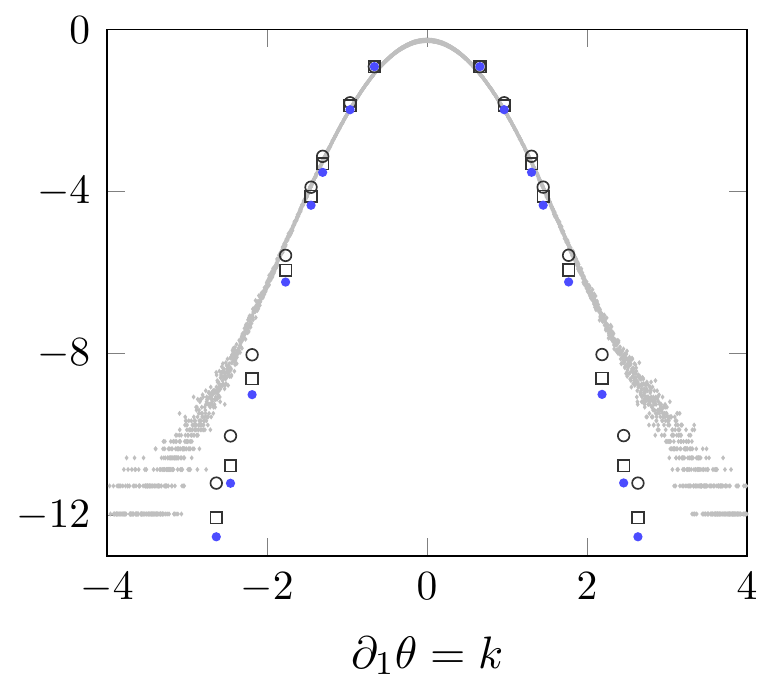}}
\caption{Log of the histogram of gradient of scalar $k$ obtained from numerical solution of SDE's (\ref{1}) and (\ref{4}) (gray dots) compared to instanton predictions: $\ln (\rho(k)) = -\Act$, parametric action Eqn. (\ref{A}) (squares); $\ln (\rho(k)) = -\bar \Act$, Jacobi/Mauperthuis action Eqn. (\ref{Abar}) (filled circles); $\ln(\rho(k)) = -\Act_{\text{eff}}$, renormalized parametric action, Eqn. (\ref{eff_act1}) with corrected parameters (\ref{gtheta_ren}) \citep{APOLINARIO} (open circles), for $g_\theta = 1.0$, $T_\theta = 0.5$, $T=1.0$ and (a) $g = 0.3$, $\textsl{variance} = 0.35$, $\textsl{kurtosis}= 3.02$; (b) $g = 0.4$, $\textsl{variance} = 0.26$, $\textsl{kurtosis}= 3.07$; (c) $g = 0.5$, $\textsl{variance} = 0.27$, $\textsl{kurtosis}= 3.17$; (d) $g=0.6$, $\textsl{variance} = 0.29$, $\textsl{kurtosis}= 3.40$;} 
\label{fig3}
\end{figure}

Different instanton prediction for logarithm scalar gradient pdfs are shown in Figure (\ref{fig3}), according to: $\ln (\rho(k)) = -\Act$, parametric action Eqn. (\ref{A}) (squares); $\ln (\rho(k)) = -\bar \Act$, Jacobi/Mauperthuis action Eqn. (\ref{Abar}) (filled circles); $\ln(\rho(k)) = -\Act_{\text{eff}}$, effective parametric action Eqn. (\ref{eff_act1}) supplemented with renormalized parameters (\ref{gtheta_ren}) \citep{APOLINARIO} (open circles).
The logarithm of pdfs calculated from the numerical solution of the SDE's (\ref{1}) and (\ref{4}) (gray dots) are also exhibited for  values of velocity gradient noise $g=0.3, 0.4, 0.5$ and $0.6$. The SDE's have been integrated with the help of a prediction-corrector algorithm \citep{WELTON1997} with step $dt = 0.01$ and final time $2\times 10^5$. It should be noted that concerning the instanton algorithm above, for noise amplitude $g=0.5$ and higher, the iterations become very unstable, failing to provide a finite result. In order to fix this we have adjusted the value of the arclength in step 4 above, decreasing the actual arclength value. For instance, for $\theta(1)>1.6$, $g=0.6$ and $E_0 = 0.4$, the arclength $L$ should be lowered down to $40 \%$. This causes a distortion in the form of instanton speed compared to the input speed. At the same time this seems to avoid instanton speed to grow indefinitely, stabilizing the iterations. More investigation on this issue is needed but will be postponed to another research.

It is clear that the method presented here, \emph{i.e.}, the parametric Hamilton's equations supplemented with noise renormalization outperforms the other approaches as can be seen by comparing the pdf's with the empirical/numerical ones Fig (\ref{fig3}). Albeit that at the cores the instanton pdfs are almost indistinguishable at the tails the instanton prediction with parametric action  and renormalization correction taken together shows a better agreement. It can be seen that for $g=0.3$ the pdfs are almost gaussian Fig. \ref{fig3}a. The kurtosis, though increases with $g$. Our instanton prediction fails to obtain the correct long tail behaviour for $g$ higher than $0.5$. It is evident that instanton approximation is not enough, and fluctuations must be taken into account. We carried out instanton fluctuation by means of noise and propagator renormalization which improves but still could not predict accurately intermittency effects as depicted in Figs. \ref{fig3}c and \ref{fig3}d.

Some interpretation can be drawn from our  results concerning geometric features of passive scalar gradient and velocity gradients in the model. We have seen that the instanton  velocity gradient configuration assigned to the major contribution to the evolution of scalar gradient is diagonal. It follows that the most probable scalar gradient fluctuation develops as a result of an underlying strain velocity field configuration. For instance, picture a fluid blob with some contaminant attached to be followed by the Lagrangian coordinates. The above reasoning means that it is more likely that the contaminant will be packed (as a result of intensification of scalar gradient) due to strain than due to local fluid rotation. Moreover, our results point out that the velocity gradient is negative along the direction of the instanton scalar gradient (Fig. \ref{fig0}a). That is, the most probable scalar gradient fluctuation points at the same direction of a compressional direction of the strain rate tensor. This finding is in line with the preferential alignment of the scalar gradient with the compressional axis of strain reported in  \cite{vedula_yeung_fox_2001,gonzalez}.

\section{Summary and perspectives}


Regarding instanton computation we have shown how to adapt the geometric arclength parametrization \cite{arclength} with constant instanton speed to account for an arbitrary intanton speed. The set of  parametric Hamilton's equations is derived exploring the geometrical aspects of the dynamical equations  invoking the Jacobi's/Mauperthuis action principle, which amounts to a variational problem searching for paths with constant Hamiltonian (energy - by abuse of notation). While the Jacobi/Mauperthuis functional agrees with the Hamilton action for vanishing energy, they numerically differ for finite energy case. From the pratical standpoint  it is often necessary to set $E \neq 0$ in order to prevent singularities in parametric Hamilton's equation. As a result, a correction accounting for finite energy and time duration of trajectory have to be considered as we have shown. 

We applied this framework to a model for Lagrangian scalar gradient dynamics, namely, the sgRFD \cite{gonzalez} in order to predict stationary pdfs and compared with pdfs derived from numerical SDE's  model. With respect to the pdfs, the finite energy correction as predicted by Jacobi/Mauperthuis action performs slightly better compared to Hamilton action. The latter underestimates the pdf. Along with that is has been shown that the parametric action yields improved results for the pdfs compared with other approximations when instanton fluctuations are considered. These fluctuations have been carried out by renormalization procedure following \citep{Moriconi_2014,APOLINARIO}.

The allowance for an arbitrary instanton speed may be useful when dealing with a phase space in the presence of multiple attractors. We speculate that it can be advantageous to control instanton speed as it navigates through or nearby attractors. This may be also used to improve the acces to large deviation paths. These questions will be left for future investigations. We also highlight that the approach outlined here can be applied to general SDEs or SPDEs. 

\section{Acknowledgments}
The author acknowledges financial support  from CEFET/RJ under project APP-CAMPI 2018 and some useful discussions with R. M. Pereira.


\bibliography{Bibliography}{}

\begin{thebibliography}{33}%
\makeatletter
\providecommand \@ifxundefined [1]{%
 \@ifx{#1\undefined}
}%
\providecommand \@ifnum [1]{%
 \ifnum #1\expandafter \@firstoftwo
 \else \expandafter \@secondoftwo
 \fi
}%
\providecommand \@ifx [1]{%
 \ifx #1\expandafter \@firstoftwo
 \else \expandafter \@secondoftwo
 \fi
}%
\providecommand \natexlab [1]{#1}%
\providecommand \enquote  [1]{``#1''}%
\providecommand \bibnamefont  [1]{#1}%
\providecommand \bibfnamefont [1]{#1}%
\providecommand \citenamefont [1]{#1}%
\providecommand \href@noop [0]{\@secondoftwo}%
\providecommand \href [0]{\begingroup \@sanitize@url \@href}%
\providecommand \@href[1]{\@@startlink{#1}\@@href}%
\providecommand \@@href[1]{\endgroup#1\@@endlink}%
\providecommand \@sanitize@url [0]{\catcode `\\12\catcode `\$12\catcode
  `\&12\catcode `\#12\catcode `\^12\catcode `\_12\catcode `\%12\relax}%
\providecommand \@@startlink[1]{}%
\providecommand \@@endlink[0]{}%
\providecommand \url  [0]{\begingroup\@sanitize@url \@url }%
\providecommand \@url [1]{\endgroup\@href {#1}{\urlprefix }}%
\providecommand \urlprefix  [0]{URL }%
\providecommand \Eprint [0]{\href }%
\providecommand \doibase [0]{http://dx.doi.org/}%
\providecommand \selectlanguage [0]{\@gobble}%
\providecommand \bibinfo  [0]{\@secondoftwo}%
\providecommand \bibfield  [0]{\@secondoftwo}%
\providecommand \translation [1]{[#1]}%
\providecommand \BibitemOpen [0]{}%
\providecommand \bibitemStop [0]{}%
\providecommand \bibitemNoStop [0]{.\EOS\space}%
\providecommand \EOS [0]{\spacefactor3000\relax}%
\providecommand \BibitemShut  [1]{\csname bibitem#1\endcsname}%
\let\auto@bib@innerbib\@empty
\bibitem [{\citenamefont {Gurarie}\ and\ \citenamefont
  {Migdal}(1996)}]{Guraire_Migdal}%
  \BibitemOpen
  \bibfield  {author} {\bibinfo {author} {\bibfnamefont {V.}~\bibnamefont
  {Gurarie}}\ and\ \bibinfo {author} {\bibfnamefont {A.}~\bibnamefont
  {Migdal}},\ }\href {\doibase 10.1103/PhysRevE.54.4908} {\bibfield  {journal}
  {\bibinfo  {journal} {Phys. Rev. E}\ }\textbf {\bibinfo {volume} {54}},\
  \bibinfo {pages} {4908} (\bibinfo {year} {1996})}\BibitemShut {NoStop}%
\bibitem [{\citenamefont {Falkovich}\ \emph {et~al.}(1996)\citenamefont
  {Falkovich}, \citenamefont {Kolokolov}, \citenamefont {Lebedev},\ and\
  \citenamefont {Migdal}}]{Falkovic_etal_96}%
  \BibitemOpen
  \bibfield  {author} {\bibinfo {author} {\bibfnamefont {G.}~\bibnamefont
  {Falkovich}}, \bibinfo {author} {\bibfnamefont {I.}~\bibnamefont
  {Kolokolov}}, \bibinfo {author} {\bibfnamefont {V.}~\bibnamefont {Lebedev}},
  \ and\ \bibinfo {author} {\bibfnamefont {A.}~\bibnamefont {Migdal}},\ }\href
  {\doibase 10.1103/PhysRevE.54.4896} {\bibfield  {journal} {\bibinfo
  {journal} {Phys. Rev. E}\ }\textbf {\bibinfo {volume} {54}},\ \bibinfo
  {pages} {4896} (\bibinfo {year} {1996})}\BibitemShut {NoStop}%
\bibitem [{\citenamefont {Chernykh}\ and\ \citenamefont
  {Stepanov}(2001)}]{Chernykh_Stepanov}%
  \BibitemOpen
  \bibfield  {author} {\bibinfo {author} {\bibfnamefont {A.~I.}\ \bibnamefont
  {Chernykh}}\ and\ \bibinfo {author} {\bibfnamefont {M.~G.}\ \bibnamefont
  {Stepanov}},\ }\href {\doibase 10.1103/PhysRevE.64.026306} {\bibfield
  {journal} {\bibinfo  {journal} {Phys. Rev. E}\ }\textbf {\bibinfo {volume}
  {64}},\ \bibinfo {pages} {026306} (\bibinfo {year} {2001})}\BibitemShut
  {NoStop}%
\bibitem [{\citenamefont {Grafke}\ \emph {et~al.}(2015)\citenamefont {Grafke},
  \citenamefont {Grauer},\ and\ \citenamefont {Sch\"aefer}}]{Grafke_2015}%
  \BibitemOpen
  \bibfield  {author} {\bibinfo {author} {\bibfnamefont {T.}~\bibnamefont
  {Grafke}}, \bibinfo {author} {\bibfnamefont {R.}~\bibnamefont {Grauer}}, \
  and\ \bibinfo {author} {\bibfnamefont {T.}~\bibnamefont {Sch\"aefer}},\
  }\href {\doibase 10.1088/1751-8113/48/33/333001} {\bibfield  {journal}
  {\bibinfo  {journal} {Journal of Physics A: Mathematical and Theoretical}\
  }\textbf {\bibinfo {volume} {48}},\ \bibinfo {pages} {333001} (\bibinfo
  {year} {2015})}\BibitemShut {NoStop}%
\bibitem [{\citenamefont {Touchette}(2009)}]{TOUCHETTE20091}%
  \BibitemOpen
  \bibfield  {author} {\bibinfo {author} {\bibfnamefont {H.}~\bibnamefont
  {Touchette}},\ }\href {\doibase
  https://doi.org/10.1016/j.physrep.2009.05.002} {\bibfield  {journal}
  {\bibinfo  {journal} {Physics Reports}\ }\textbf {\bibinfo {volume} {478}},\
  \bibinfo {pages} {1 } (\bibinfo {year} {2009})}\BibitemShut {NoStop}%
\bibitem [{\citenamefont {Rolland}\ \emph {et~al.}(2016)\citenamefont
  {Rolland}, \citenamefont {Bouchet},\ and\ \citenamefont
  {Simonnet}}]{Bouchet_etal_2016_GL}%
  \BibitemOpen
  \bibfield  {author} {\bibinfo {author} {\bibfnamefont {J.}~\bibnamefont
  {Rolland}}, \bibinfo {author} {\bibfnamefont {F.}~\bibnamefont {Bouchet}}, \
  and\ \bibinfo {author} {\bibfnamefont {E.}~\bibnamefont {Simonnet}},\ }\href
  {http://search-ebscohost-com.ez108.periodicos.capes.gov.br/login.aspx?direct=true&db=aph&AN=112192013&lang=pt-br&site=ehost-live&authtype=ip,cookie,uid}
  {\bibfield  {journal} {\bibinfo  {journal} {Journal of Statistical Physics}\
  }\textbf {\bibinfo {volume} {162}},\ \bibinfo {pages} {277 } (\bibinfo {year}
  {2016})}\BibitemShut {NoStop}%
\bibitem [{\citenamefont {Giardin\`a}\ \emph {et~al.}(2006)\citenamefont
  {Giardin\`a}, \citenamefont {Kurchan},\ and\ \citenamefont
  {Peliti}}]{Giardina_etal_PRL2006}%
  \BibitemOpen
  \bibfield  {author} {\bibinfo {author} {\bibfnamefont {C.}~\bibnamefont
  {Giardin\`a}}, \bibinfo {author} {\bibfnamefont {J.}~\bibnamefont {Kurchan}},
  \ and\ \bibinfo {author} {\bibfnamefont {L.}~\bibnamefont {Peliti}},\ }\href
  {\doibase 10.1103/PhysRevLett.96.120603} {\bibfield  {journal} {\bibinfo
  {journal} {Phys. Rev. Lett.}\ }\textbf {\bibinfo {volume} {96}},\ \bibinfo
  {pages} {120603} (\bibinfo {year} {2006})}\BibitemShut {NoStop}%
\bibitem [{\citenamefont {Lestang}\ \emph {et~al.}(2018)\citenamefont
  {Lestang}, \citenamefont {Ragone}, \citenamefont {Br{\'{e}}hier},
  \citenamefont {Herbert},\ and\ \citenamefont {Bouchet}}]{Lestang_2018}%
  \BibitemOpen
  \bibfield  {author} {\bibinfo {author} {\bibfnamefont {T.}~\bibnamefont
  {Lestang}}, \bibinfo {author} {\bibfnamefont {F.}~\bibnamefont {Ragone}},
  \bibinfo {author} {\bibfnamefont {C.-E.}\ \bibnamefont {Br{\'{e}}hier}},
  \bibinfo {author} {\bibfnamefont {C.}~\bibnamefont {Herbert}}, \ and\
  \bibinfo {author} {\bibfnamefont {F.}~\bibnamefont {Bouchet}},\ }\href
  {\doibase 10.1088/1742-5468/aab856} {\bibfield  {journal} {\bibinfo
  {journal} {Journal of Statistical Mechanics: Theory and Experiment}\ }\textbf
  {\bibinfo {volume} {2018}},\ \bibinfo {pages} {043213} (\bibinfo {year}
  {2018})}\BibitemShut {NoStop}%
\bibitem [{\citenamefont {E}\ \emph {et~al.}(2004)\citenamefont {E},
  \citenamefont {Ren},\ and\ \citenamefont {Vanden-Eijnden}}]{mam}%
  \BibitemOpen
  \bibfield  {author} {\bibinfo {author} {\bibfnamefont {W.}~\bibnamefont {E}},
  \bibinfo {author} {\bibfnamefont {W.}~\bibnamefont {Ren}}, \ and\ \bibinfo
  {author} {\bibfnamefont {E.}~\bibnamefont {Vanden-Eijnden}},\ }\href@noop {}
  {\bibfield  {journal} {\bibinfo  {journal} {Commun. Pure Appl. Math.}\
  }\textbf {\bibinfo {volume} {57}},\ \bibinfo {pages} {637} (\bibinfo {year}
  {2004})}\BibitemShut {NoStop}%
\bibitem [{\citenamefont {E}\ \emph {et~al.}(2002)\citenamefont {E},
  \citenamefont {Ren},\ and\ \citenamefont {Vanden-Eijnden}}]{string}%
  \BibitemOpen
  \bibfield  {author} {\bibinfo {author} {\bibfnamefont {W.}~\bibnamefont {E}},
  \bibinfo {author} {\bibfnamefont {W.}~\bibnamefont {Ren}}, \ and\ \bibinfo
  {author} {\bibfnamefont {E.}~\bibnamefont {Vanden-Eijnden}},\ }\href
  {\doibase 10.1103/PhysRevB.66.052301} {\bibfield  {journal} {\bibinfo
  {journal} {Phys. Rev. B}\ }\textbf {\bibinfo {volume} {66}},\ \bibinfo
  {pages} {052301} (\bibinfo {year} {2002})}\BibitemShut {NoStop}%
\bibitem [{\citenamefont {E}\ \emph {et~al.}(2007)\citenamefont {E},
  \citenamefont {Ren},\ and\ \citenamefont {Vanden-Eijnden}}]{string2}%
  \BibitemOpen
  \bibfield  {author} {\bibinfo {author} {\bibfnamefont {W.}~\bibnamefont {E}},
  \bibinfo {author} {\bibfnamefont {W.}~\bibnamefont {Ren}}, \ and\ \bibinfo
  {author} {\bibfnamefont {E.}~\bibnamefont {Vanden-Eijnden}},\ }\href
  {\doibase 10.1063/1.2720838} {\bibfield  {journal} {\bibinfo  {journal} {The
  Journal of chemical physics}\ }\textbf {\bibinfo {volume} {126}},\ \bibinfo
  {pages} {164103} (\bibinfo {year} {2007})}\BibitemShut {NoStop}%
\bibitem [{\citenamefont {Heymann}\ and\ \citenamefont
  {Vanden-Eijnden}(2008{\natexlab{a}})}]{pathways}%
  \BibitemOpen
  \bibfield  {author} {\bibinfo {author} {\bibfnamefont {M.}~\bibnamefont
  {Heymann}}\ and\ \bibinfo {author} {\bibfnamefont {E.}~\bibnamefont
  {Vanden-Eijnden}},\ }\href {\doibase 10.1103/PhysRevLett.100.140601}
  {\bibfield  {journal} {\bibinfo  {journal} {Phys. Rev. Lett.}\ }\textbf
  {\bibinfo {volume} {100}},\ \bibinfo {pages} {140601} (\bibinfo {year}
  {2008}{\natexlab{a}})}\BibitemShut {NoStop}%
\bibitem [{\citenamefont {Heymann}\ and\ \citenamefont
  {Vanden-Eijnden}(2008{\natexlab{b}})}]{gmam}%
  \BibitemOpen
  \bibfield  {author} {\bibinfo {author} {\bibfnamefont {M.}~\bibnamefont
  {Heymann}}\ and\ \bibinfo {author} {\bibfnamefont {E.}~\bibnamefont
  {Vanden-Eijnden}},\ }\href {\doibase 10.1002/cpa.20238} {\bibfield  {journal}
  {\bibinfo  {journal} {Communications on Pure and Applied Mathematics}\
  }\textbf {\bibinfo {volume} {61}},\ \bibinfo {pages} {1052 } (\bibinfo {year}
  {2008}{\natexlab{b}})}\BibitemShut {NoStop}%
\bibitem [{\citenamefont {Grafke}\ \emph {et~al.}(2013)\citenamefont {Grafke},
  \citenamefont {Grauer}, \citenamefont {Sch\"afer},\ and\ \citenamefont
  {Vanden-Eijnden}}]{arclength}%
  \BibitemOpen
  \bibfield  {author} {\bibinfo {author} {\bibfnamefont {T.}~\bibnamefont
  {Grafke}}, \bibinfo {author} {\bibfnamefont {R.}~\bibnamefont {Grauer}},
  \bibinfo {author} {\bibfnamefont {T.}~\bibnamefont {Sch\"afer}}, \ and\
  \bibinfo {author} {\bibfnamefont {E.}~\bibnamefont {Vanden-Eijnden}},\ }\href
  {\doibase 10.1137/130939158} {\bibfield  {journal} {\bibinfo  {journal}
  {Multiscale Model. Simul.}\ }\textbf {\bibinfo {volume} {12}},\ \bibinfo
  {pages} {566} (\bibinfo {year} {2013})}\BibitemShut {NoStop}%
\bibitem [{\citenamefont {Lanczos}(1986)}]{Lanczos}%
  \BibitemOpen
  \bibfield  {author} {\bibinfo {author} {\bibfnamefont {C.}~\bibnamefont
  {Lanczos}},\ }\href@noop {} {\emph {\bibinfo {title} {The Variational
  Principles of Mechanics}}},\ \bibinfo {edition} {4th}\ ed.\ (\bibinfo
  {publisher} {Dover Publications},\ \bibinfo {address} {New York},\ \bibinfo
  {year} {1986})\BibitemShut {NoStop}%
\bibitem [{\citenamefont {Gonzalez}(2009)}]{gonzalez}%
  \BibitemOpen
  \bibfield  {author} {\bibinfo {author} {\bibfnamefont {M.}~\bibnamefont
  {Gonzalez}},\ }\href@noop {} {\bibfield  {journal} {\bibinfo  {journal}
  {Phys. Flu.}\ }\textbf {\bibinfo {volume} {21}},\ \bibinfo {pages} {055104}
  (\bibinfo {year} {2009})}\BibitemShut {NoStop}%
\bibitem [{\citenamefont {Hater}\ \emph {et~al.}(2011)\citenamefont {Hater},
  \citenamefont {Homann},\ and\ \citenamefont {Grauer}}]{rfd_magnetic}%
  \BibitemOpen
  \bibfield  {author} {\bibinfo {author} {\bibfnamefont {T.}~\bibnamefont
  {Hater}}, \bibinfo {author} {\bibfnamefont {H.}~\bibnamefont {Homann}}, \
  and\ \bibinfo {author} {\bibfnamefont {R.}~\bibnamefont {Grauer}},\ }\href
  {\doibase 10.1103/PhysRevE.83.017302} {\bibfield  {journal} {\bibinfo
  {journal} {Phys. Rev. E}\ }\textbf {\bibinfo {volume} {83}},\ \bibinfo
  {pages} {017302} (\bibinfo {year} {2011})}\BibitemShut {NoStop}%
\bibitem [{\citenamefont {Moriconi}\ \emph {et~al.}(2014)\citenamefont
  {Moriconi}, \citenamefont {Pereira},\ and\ \citenamefont
  {Grigorio}}]{Moriconi_2014}%
  \BibitemOpen
  \bibfield  {author} {\bibinfo {author} {\bibfnamefont {L.}~\bibnamefont
  {Moriconi}}, \bibinfo {author} {\bibfnamefont {R.~M.}\ \bibnamefont
  {Pereira}}, \ and\ \bibinfo {author} {\bibfnamefont {L.~S.}\ \bibnamefont
  {Grigorio}},\ }\href {\doibase 10.1088/1742-5468/2014/10/p10015} {\bibfield
  {journal} {\bibinfo  {journal} {Journal of Statistical Mechanics: Theory and
  Experiment}\ }\textbf {\bibinfo {volume} {2014}},\ \bibinfo {pages} {P10015}
  (\bibinfo {year} {2014})}\BibitemShut {NoStop}%
\bibitem [{\citenamefont {Apolinário}\ \emph {et~al.}(2019)\citenamefont
  {Apolinário}, \citenamefont {Moriconi},\ and\ \citenamefont
  {Pereira}}]{APOLINARIO}%
  \BibitemOpen
  \bibfield  {author} {\bibinfo {author} {\bibfnamefont {G.}~\bibnamefont
  {Apolinário}}, \bibinfo {author} {\bibfnamefont {L.}~\bibnamefont
  {Moriconi}}, \ and\ \bibinfo {author} {\bibfnamefont {R.}~\bibnamefont
  {Pereira}},\ }\href {\doibase https://doi.org/10.1016/j.physa.2018.09.102}
  {\bibfield  {journal} {\bibinfo  {journal} {Physica A: Statistical Mechanics
  and its Applications}\ }\textbf {\bibinfo {volume} {514}},\ \bibinfo {pages}
  {741 } (\bibinfo {year} {2019})}\BibitemShut {NoStop}%
\bibitem [{Notea()}]{Notea}%
  \BibitemOpen
  \bibinfo {note} {Can be interpreted as a Legendre transform.}\BibitemShut
  {Stop}%
\bibitem [{\citenamefont {Gray}(2009)}]{Scholarpedia_leastaction}%
  \BibitemOpen
  \bibfield  {author} {\bibinfo {author} {\bibfnamefont {C.~G.}\ \bibnamefont
  {Gray}},\ }\href {\doibase 10.4249/scholarpedia.8291} {\bibfield  {journal}
  {\bibinfo  {journal} {Scholarpedia}\ }\textbf {\bibinfo {volume} {4}},\
  \bibinfo {pages} {8291} (\bibinfo {year} {2009})},\ \bibinfo {note} {revision
  \#150617}\BibitemShut {NoStop}%
\bibitem [{\citenamefont {Chevillard}\ and\ \citenamefont
  {Meneveau}(2006)}]{RFD}%
  \BibitemOpen
  \bibfield  {author} {\bibinfo {author} {\bibfnamefont {L.}~\bibnamefont
  {Chevillard}}\ and\ \bibinfo {author} {\bibfnamefont {C.}~\bibnamefont
  {Meneveau}},\ }\href {\doibase 10.1103/PhysRevLett.97.174501} {\bibfield
  {journal} {\bibinfo  {journal} {Phys. Rev. Lett.}\ }\textbf {\bibinfo
  {volume} {97}},\ \bibinfo {pages} {174501} (\bibinfo {year}
  {2006})}\BibitemShut {NoStop}%
\bibitem [{\citenamefont {Honeycutt}(1992)}]{Honeycutt_92}%
  \BibitemOpen
  \bibfield  {author} {\bibinfo {author} {\bibfnamefont {R.~L.}\ \bibnamefont
  {Honeycutt}},\ }\href {\doibase 10.1103/PhysRevA.45.600} {\bibfield
  {journal} {\bibinfo  {journal} {Phys. Rev. A}\ }\textbf {\bibinfo {volume}
  {45}},\ \bibinfo {pages} {600} (\bibinfo {year} {1992})}\BibitemShut
  {NoStop}%
\bibitem [{\citenamefont {Welton}\ and\ \citenamefont
  {Pope}(1997)}]{WELTON1997}%
  \BibitemOpen
  \bibfield  {author} {\bibinfo {author} {\bibfnamefont {W.~C.}\ \bibnamefont
  {Welton}}\ and\ \bibinfo {author} {\bibfnamefont {S.~B.}\ \bibnamefont
  {Pope}},\ }\href {\doibase https://doi.org/10.1006/jcph.1997.5680} {\bibfield
   {journal} {\bibinfo  {journal} {Journal of Computational Physics}\ }\textbf
  {\bibinfo {volume} {134}},\ \bibinfo {pages} {150 } (\bibinfo {year}
  {1997})}\BibitemShut {NoStop}%
\bibitem [{\citenamefont {Martin}\ \emph {et~al.}(1973)\citenamefont {Martin},
  \citenamefont {Siggia},\ and\ \citenamefont {Rose}}]{PhysRevA.8.423}%
  \BibitemOpen
  \bibfield  {author} {\bibinfo {author} {\bibfnamefont {P.~C.}\ \bibnamefont
  {Martin}}, \bibinfo {author} {\bibfnamefont {E.~D.}\ \bibnamefont {Siggia}},
  \ and\ \bibinfo {author} {\bibfnamefont {H.~A.}\ \bibnamefont {Rose}},\
  }\href {\doibase 10.1103/PhysRevA.8.423} {\bibfield  {journal} {\bibinfo
  {journal} {Phys. Rev. A}\ }\textbf {\bibinfo {volume} {8}},\ \bibinfo {pages}
  {423} (\bibinfo {year} {1973})}\BibitemShut {NoStop}%
\bibitem [{\citenamefont {Janssen}(1976)}]{Janssen1976}%
  \BibitemOpen
  \bibfield  {author} {\bibinfo {author} {\bibfnamefont {H.-K.}\ \bibnamefont
  {Janssen}},\ }\href {\doibase 10.1007/BF01316547} {\bibfield  {journal}
  {\bibinfo  {journal} {Zeitschrift f{\"u}r Physik B Condensed Matter}\
  }\textbf {\bibinfo {volume} {23}},\ \bibinfo {pages} {377} (\bibinfo {year}
  {1976})}\BibitemShut {NoStop}%
\bibitem [{\citenamefont {de~Dominicis}(1976)}]{Dominicis_76}%
  \BibitemOpen
  \bibfield  {author} {\bibinfo {author} {\bibfnamefont {C.}~\bibnamefont
  {de~Dominicis}},\ }\href@noop {} {\bibfield  {journal} {\bibinfo  {journal}
  {J. Physique C}\ }\textbf {\bibinfo {volume} {37}},\ \bibinfo {pages} {247}
  (\bibinfo {year} {1976})}\BibitemShut {NoStop}%
\bibitem [{\citenamefont {McComb}(2009)}]{mccomb}%
  \BibitemOpen
  \bibfield  {author} {\bibinfo {author} {\bibfnamefont {W.~D.}\ \bibnamefont
  {McComb}},\ }\href@noop {} {\emph {\bibinfo {title} {Renormalization Methods
  - a guide for beginners}}},\ \bibinfo {edition} {1st}\ ed.\ (\bibinfo
  {publisher} {Claredom Press - Oxford},\ \bibinfo {address} {New York},\
  \bibinfo {year} {2009})\BibitemShut {NoStop}%
\bibitem [{\citenamefont {Barabasi}\ and\ \citenamefont
  {Stanley}(1995)}]{barabasi-stanley}%
  \BibitemOpen
  \bibfield  {author} {\bibinfo {author} {\bibfnamefont {A.~L.}\ \bibnamefont
  {Barabasi}}\ and\ \bibinfo {author} {\bibfnamefont {E.}~\bibnamefont
  {Stanley}},\ }\href@noop {} {\emph {\bibinfo {title} {Fractal Concepts in
  Surface Growth}}},\ \bibinfo {edition} {1st}\ ed.\ (\bibinfo  {publisher}
  {Cambridge University Press},\ \bibinfo {address} {New York},\ \bibinfo
  {year} {1995})\BibitemShut {NoStop}%
\bibitem [{\citenamefont {Chow}\ and\ \citenamefont
  {Buice}(2010)}]{Chow2010PathIM}%
  \BibitemOpen
  \bibfield  {author} {\bibinfo {author} {\bibfnamefont {C.}~\bibnamefont
  {Chow}}\ and\ \bibinfo {author} {\bibfnamefont {M.~A.}\ \bibnamefont
  {Buice}},\ }\href@noop {} {\bibfield  {journal} {\bibinfo  {journal} {Journal
  of Mathematical Neuroscience}\ }\textbf {\bibinfo {volume} {5}} (\bibinfo
  {year} {2010})}\BibitemShut {NoStop}%
\bibitem [{\citenamefont {Wang}(2013)}]{MathGR}%
  \BibitemOpen
  \bibfield  {author} {\bibinfo {author} {\bibfnamefont {Y.}~\bibnamefont
  {Wang}},\ }\href {http://arxiv.org/abs/1306.1295} {\bibfield  {journal}
  {\bibinfo  {journal} {CoRR}\ }\textbf {\bibinfo {volume} {abs/1306.1295}}
  (\bibinfo {year} {2013})},\ \Eprint {http://arxiv.org/abs/1306.1295}
  {arXiv:1306.1295} \BibitemShut {NoStop}%
\bibitem [{\citenamefont {Grigorio}\ \emph {et~al.}(2017)\citenamefont
  {Grigorio}, \citenamefont {Bouchet}, \citenamefont {Pereira},\ and\
  \citenamefont {Chevillard}}]{Grigorio_2017}%
  \BibitemOpen
  \bibfield  {author} {\bibinfo {author} {\bibfnamefont {L.~S.}\ \bibnamefont
  {Grigorio}}, \bibinfo {author} {\bibfnamefont {F.}~\bibnamefont {Bouchet}},
  \bibinfo {author} {\bibfnamefont {R.~M.}\ \bibnamefont {Pereira}}, \ and\
  \bibinfo {author} {\bibfnamefont {L.}~\bibnamefont {Chevillard}},\ }\href
  {\doibase 10.1088/1751-8121/aa51a3} {\bibfield  {journal} {\bibinfo
  {journal} {Journal of Physics A: Mathematical and Theoretical}\ }\textbf
  {\bibinfo {volume} {50}},\ \bibinfo {pages} {055501} (\bibinfo {year}
  {2017})}\BibitemShut {NoStop}%
\bibitem [{\citenamefont {Vedula}\ \emph {et~al.}(2001)\citenamefont {Vedula},
  \citenamefont {Yeung},\ and\ \citenamefont {Fox}}]{vedula_yeung_fox_2001}%
  \BibitemOpen
  \bibfield  {author} {\bibinfo {author} {\bibfnamefont {P.}~\bibnamefont
  {Vedula}}, \bibinfo {author} {\bibfnamefont {P.~K.}\ \bibnamefont {Yeung}}, \
  and\ \bibinfo {author} {\bibfnamefont {R.~O.}\ \bibnamefont {Fox}},\ }\href
  {\doibase 10.1017/S0022112000003207} {\bibfield  {journal} {\bibinfo
  {journal} {Journal of Fluid Mechanics}\ }\textbf {\bibinfo {volume} {433}},\
  \bibinfo {pages} {29–60} (\bibinfo {year} {2001})}\BibitemShut {NoStop}%
\end{thebibliography}%

\end{document}